\documentclass[review]{elsarticle}

\usepackage{lineno,hyperref}
\usepackage{amsmath}
\modulolinenumbers[5]

\journal{Radiation Physics and Chemistry}

\begin{document}

\begin{frontmatter}

\title{A hybrid model for estimation of pore size from ortho-positronium lifetimes in porous materials}


\author[CNT,Dubna]{L. Anh Tuyen\corref{mycorrespondingauthor}}
\ead{tuyenluuanh@gmail.com}
\address[CNT]{Center for Nuclear Techniques, Vietnam Atomic Energy Institute, 217 Nguyen Trai, District 1, Ho Chi Minh City, Vietnam}
\address[Dubna]{Joint Institute for Nuclear Research, 6 Joliot Curie, Dubna, Russia}
\author[IFAS]{T. Dong Xuan}
\address[IFAS]{Institute of Fundamental and Applied Sciences, Duy Tan University, Ho Chi Minh City, Vietnam}
\author[DTU,UPEN]{H. A. Tuan Kiet}
\address[DTU]{Institute of Research and Development, Duy Tan University, Danang City, Vietnam}
\address[UPEN]{Graduate School of Education, University of Pennsylvania, Philadelphia, Pennsylvania 19104, USA}
\author[UTE]{L. Chi Cuong}
\address[UTE]{University of Technical Education, 1 Vo Van Ngan, Thu Duc District, Ho Chi Minh City, Vietnam}
\author[CNT]{P. Trong Phuc}
\author[HCMU]{T. Duy Tap}
\address[HCMU]{University of Science, Vietnam National University Ho Chi Minh City, 227 Nguyen Van Cu, District 5, Ho Chi Minh City, Vietnam}
\author[IFAS]{Dinh-Van Phuc}
\author[CNT,HCMU]{L. Ly Nguyen}
\author[CNT,HCMU]{N. T. Ngoc Hue}
\author[CNT]{P. Thi Hue}
\author[CNT,HCMU]{L. Thai Son}
\author[CNT]{D. Van Hoang}
\author[CNT]{N. Hoang Long}
\author[IFAS]{N. Quang Hung\corref{mycorrespondingauthor}}
\ead{nguyenquanghung5@duytan.edu.vn}
\cortext[mycorrespondingauthor]{Corresponding Author}

\begin{abstract}
The present paper proposes a novel model for estimating the free-volume size of the porous materials based on the analysis of the experimental ortho-positronium ($o$-Ps) lifetimes collecting within more than four decades. The model is derived by combining the semi-classical (SE) physics model, which works in the region of large pores (pore size $R >$ 1 nm), with the conventional Tao-Eldrup (TE) model, which is applicable only for the small-pore region ($R <$ 1 nm). Thus, the resulting model, called the hybrid (HYB) model, is able to smoothly connect the $o$-Ps lifetimes in the two regions of the pore. Moreover, by introducing the $o$-Ps diffusion probability parameter ($D$), the HYB model has reproduced quite well the experimental $o$-Ps lifetimes in the whole region of pore sizes. It is even in a better agreement with the experimental data than the most up-to-date rectangular TE (RTE) and Tokyo models. In particular, by adjusting the value of $D$, the HYB model can also describe very well the two defined sets of experimental $o$-Ps lifetime in the pores with spherical and channel geometries. The merit of the present model, in comparison with the previously proposed ones, is that it is applicable for the pore size in the universal range of $0.2 - 400$ nm for most of porous materials with different geometries. 
\end{abstract}


\end{frontmatter}


\section{Introduction}
Using and controlling the reciprocal influences between porosity, crystal structure, lattice defects, surface property, molecular transport, and reactions (chemical, catalytic, ion-exchange, ect) in functional materials are keys to various technologies for sustainable development of the world \cite{Davis02}. Nanostructured materials such as silica, zeolite, metal-organic framework materials, etc., contain a complex matrix of pore structures, which plays a fundamental role to their applications in environmental treatments, industrial catalysis, energy storage, etc. In the structural research of above materials, the applicable limitations of adsorbed/desorbed methods at the nanometer scale lead to participation of the nuclear techniques such as the small angle neutron scattering (SANS) \cite{SANS}, small angle X-ray scattering (SAXS) \cite{SAXS}, positron annihilation lifetime spectroscopy (PALS) \cite{PALS, Gidley76}, ect. Among these techniques, PALS has been commonly and widely applied for the structure study of various materials and matters in many decades, including solids \cite{Kaha63}, molecular substances \cite{Tao72}, electron liquid \cite{Arp79} and gas \cite{Drum11}, solid pivalic acid \cite{Eldrup81}, thin films \cite{Dull01}, semiconductors \cite{Tuo13}, zeolites \cite{Tuyen15,Tuyen17}, etc. One of the most obvious applications of PALS arises from the fact that in materials, the structural traps such as micro/mesopores (channels, cavities or cages) and defects (vacancies, voids, etc.) could affect the local electron distributions and consequently lead to a significant change of the positron lifetimes \cite{Dull01,Kaj07}. As a result, the localized $o$-Ps lifetime is sensitive to the trap size, that is, the larger the trap is, the longer lifetime of the localized $o$-Ps can be achieved \cite{Tuyen17}. For the spherical pores, the first and simplest model describing the correlation of $o$-Ps lifetime ($1/\lambda$) and pore radius ($R$) via the pick-off annihilation was proposed long time ago by Tao and Eldrup (TE model) by using a spherically symmetric infinite potential well \cite{Tao72,Eldrup81}. The TE model, which was derived based on the quantum mechanics, considers the $o$-Ps in the pore as a single particle, whose mass is twice the electron mass, in a spherical potential well of radius $R$ (pore radius). Thus, this model is applicable only for estimating the pores with very small size $R$ of less than 1 nm. In other words, this model is not applicable for $R > 1$ nm \cite{Gow02}.

Many attempts have been made since last three decades in oder to extend the TE model to the region of larger pores. The first attempt was made by extending the TE model (also called the ETE model) to include some thermally excited states in the spherical and infinitely long cylindrical wells using the Bessel functions \cite{Gow98}. However, this approach becomes computationally difficult once the order of the spherical Bessel functions is high. A year later, a classical model (CM) was proposed by Gidley {\it et al} \cite{Gidley99}. This model is indeed an extension of the quantum mechanical TE model to the classical large-pore limit by assuming the rectangular geometry of the pores. Although this model is adequately accounted for $R$ in the range of 0.1 - 600 nm, it always overestimates the experimental data in the small pore-size regime ($R < 1$ nm). At the same year, Ito {\it et al.} proposed the modified TE model (called Tokyo model) based on a phenomenological assumption that the $o$-Ps exhibits more like a quantum particle, bouncing back and forth between the energy barriers as the potential well or pore size is extended \cite{Ito99}. With this assumption, they have derived a simple analytical relation between the pore size and the $o$-Ps lifetime but the results obtained agree with the experimental data only at 1 nm  $< R <$ 30 nm \cite{Dull01}. The latest updated model, called the rectangular Tao-Enldrup model (RTE), was proposed by Dull {\it et al} \cite{Dull01}. The RTE was derived by switching to the rectangular pore geometry instead of the spherical one as in the conventional TE model and then taking into account the temperature effect (within the Boltzmann statistics) of the $o$-Ps lifetime. Since then, this model has been commonly used because it is applicable for pores of any size at any sample temperature. However, the later tests by Thraenert {\it et al} \cite{Thrae07} and Zaleski {\it et al} \cite{Zaleski12} have shown that the RTE agrees with the experimental data at $R > 2-3$ nm and around the room temperature ($T \approx 300$ K) only. At $R$ below $2-3$ nm, the RTE predicts the lifetimes of about 20$\%$ lower than those predicted by the ETE as well as the experimental data. In addition, at low temperatures, for small pores ($R < $3 nm), this model can not explain the increase of the experimental lifetime, whereas for larger pores ($R >$ 5nm), it shows a contrary behavior with the data. Nevertheless, the RTE has still considered as the most up-to-date extension of the TE model and no further updated model has been proposed in almost two decades. It is therefore highly desirable to develop a new model, which overcomes the shortcomings of all the above approaches. 

In fact, we have recently proposed a semi-empirical model based on the semi-classical (SE) approach, which was used to determine the $o$-Ps radial probability function for the large pores, and weighting with the TE model, which is applicable for the small pores \cite{Thanh08}. This semi-empirical model offers a better agreement with the experimental data than the ETE and MTE but only in the region of pore size from 1 nm to around 30 nm. Moreover, this approach does not take into account either the geometry of the pore or the effect of temperature. In the present paper, we report a new model for calculating the universal free volume size in porous materials by combining the quantum (TE) with the semi-classical (SE) models, taking into account different pore geometries. To determine the model parameters, the experimental data of the $o$-Ps lifetimes in porous materials reported and updated over than past four decades are used. In addition, the correctness of the model is verified by using the $o$-Ps lifetimes in the spherical and channel pores of porous materials.

\section{Theoretical models of $o$-Ps lifetime and pore-size estimations} 
\subsection{The Tao-Eldrup model for small pores}
\label{TEM}
The Tao-Eldrup model, which was the first model describing the relation between the $o$-Ps lifetime and pore radius in porous materials, assumes that the $o$-Ps in the pore can be treated as a single quantum particle moving in a spherically symmetric infinite potential well of radius $R$ (pore radius) \cite{Tao72,Eldrup81}. The pick-off annihilation rate of the trapped $o$-Ps in the pore of radius $R+\Delta R$ in this case is given by 
\begin{equation}
\lambda_{TE} = \lambda_A\left[1-\frac{R}{R+\Delta R}+\frac{1}{2\pi}\sin\bigg(\frac{2\pi R}{R+\Delta R} \bigg) \right] = \lambda_A f_{TE}(R)~, \label{TE}
\end{equation}
where $f_{TE}(R) = 1-\frac{R}{R+\Delta R}+\frac{1}{2\pi}\sin\bigg(\frac{2\pi R}{R+\Delta R} \bigg)$ is the pore-size correlation function, whereas $\lambda_A = (\lambda_S + 3\lambda_T)/4 \approx 2$ ns$^{-1}$ is the spin-averaged vacuum annihilation rate with $\lambda_S$ and $\lambda_T$ being, respectively, the intrinsic annihilation rates of spin-singlet Ps ($p$-Ps) and $o$-Ps in the vacuum. The value of the parameter $\Delta R$, which is the thickness of the virtual electron layer, was empirically obtained to be 0.166 nm. The lifetime of the $o$-Ps is simply calculated from equation (\ref{TE}) as $\tau_{TE}(R) = 1/\lambda_{TE}(R)$. The TE model describes very well the porous materials with pore size less than 1 nm. However, when the pore size is larger than 1 nm, the $o$-Ps lifetime calculated from the TE model deviates from the experimental values. The reason is that the TE model neglected the annihilation rate of the $o$-Ps lifetime in vacuum $\lambda_T$, which is equal to $\sim$ 1/142 ns \cite{Dull01}. 

\subsection{The models for large pores}
The simplest way to extend the TE model for the estimation of large pores is to modify Eq. (\ref{TE}) by adding $\lambda_T$ to its right-hand side. This semi-phenomenological model, which was proposed by a group of the University of Tokyo and thus called the Tokyo model \cite{Ito99, Dull01, Thrae07}, divided the pore radius into two regions according to a critical radius $R_a$, namely
\begin{equation}
\lambda_{Tokyo}(R) =
\begin{cases}
\lambda_{TE}(R) + \lambda_T & {\rm for} ~ R < R_a ~, \\
\lambda_{TE}(R_a)\left[1-\bigg(\frac{R-R_a}{R+\Delta R} \bigg)\right]^b + \lambda_T & {\rm for} ~ R \geq R_a ~,
\end{cases}
\label{Tokyo}
\end{equation}
where $R_a$ and $b$ are two free parameters, whose values are empirically determined via the fitting to various experimental data \cite{Thrae07}. Although the Tokyo model is able to treat the $o$-Ps lifetime in the large pores, it overestimates the experimental data when the pore size $R$ is smaller than 1 nm and higher than 30 nm (see e.g., the dotted lines in Fig. 6 of Ref. \cite{Dull01}). 

The above drawback of the Tokyo model can be solved by using the rectangular TE (RTE) model \cite{Dull01}. The RTE model was proposed by switching the pores from the spherical to rectangular geometries. By using this rectangular geometry, the RTE model has avoided to use the complicated Bessel functions as in the spherical case. The RTE equation is obtained by solving the Schrodinger equation for the $o$-Ps in an infinite rectangular well with side length of $a, b$, and $c$ in the $x, y$, and $z$ directions. The final pick-off annihilation rate of the $o$-Ps is calculated from the following equation
\begin{equation}
\lambda_{RTE}(a, b, c, T) = \lambda_A - \frac{\lambda_S - \lambda_T}{4}F(a, \delta, T)F(b, \delta, T)F(c, \delta, T) ~,
\label{RTE}
\end{equation} 
where 
\begin{equation}
F(x, \delta, T) = 1 - \frac{2\delta}{x}+\frac{\sum_{i=1}^{\infty}\frac{1}{i\pi}\sin\left(\frac{2i\pi\delta}{x}\right)\exp\bigg(-\frac{\beta i^2}{kT x^2}\bigg)}{\sum_{i=1}^\infty\exp\bigg(-\frac{\beta i^2}{kT x^2}\bigg)} ~,
\end{equation}
where $T$ is temperature, $\beta = h^2/16$ m $= 0.188$ eVnm$^2$, and $\delta$ is a free parameter, which is analogous to $\Delta R$ in Eq. (\ref{TE}) for the TE model. The RTE model contains three limiting cases, namely thin one-dimensional (1D) sheet-like pores, square infinitely long two-dimensional (2D) channel-like pores, and compact three-dimensional (3D) cubic pores. Amongst these cases, the 3D cubic case with the cube width of $a$, which relates with the pore radius $R$ via $a = 2(R+\Delta R)$, was popularly used. The value of $\delta=$ 0.18 nm was chosen so that $\lambda_{RTE}(R, T=0)$ agrees with $\lambda_{TE}(R)$ in the region in which the TE model offers the most accurate results, namely the region of $R <$ 1 nm.

\subsection{The semiclassical model}
\begin{figure}
\begin{center}
       \includegraphics[scale=0.4]{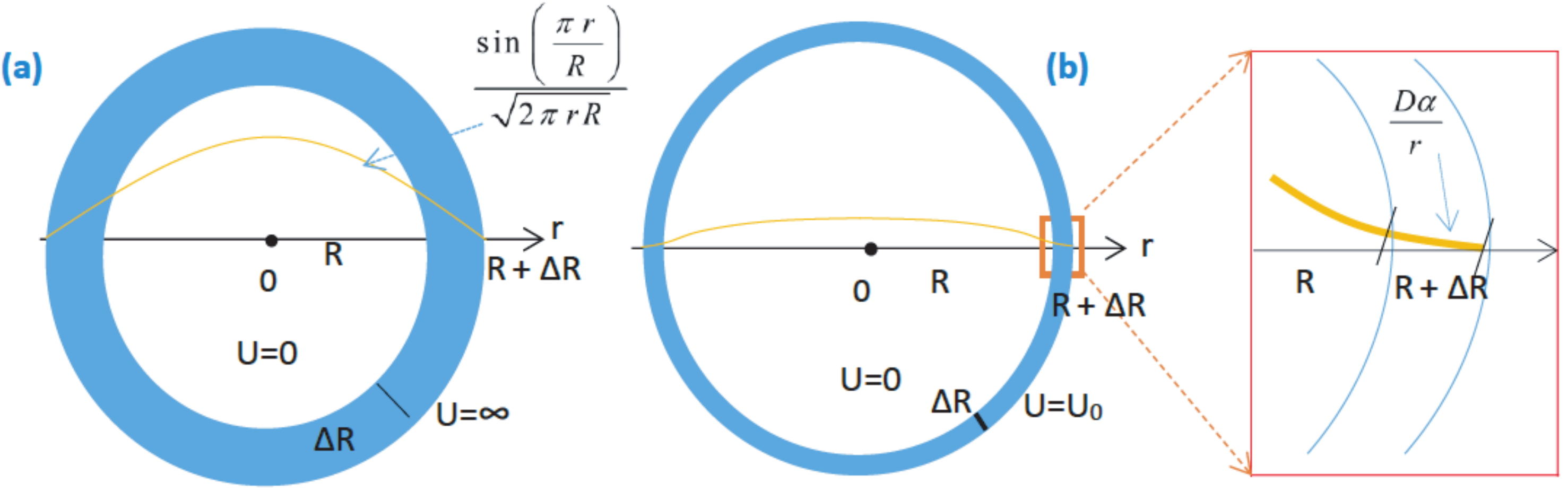}
       \caption{Illustrations of (a) the TE model for the $o$-Ps in small pores within the spherically symmetric infinite potential well. (b) the semi-classical (SE) model for the $o$-Ps in large pores within the spherically finite potential well.
        \label{fig1}}
\end{center}
\end{figure}
The semiclassical model (SE) model was first introduced in Ref. \cite{Thanh08}. This model was derived based on an assumption that in the region of small pores ($R <$ 1 nm), the $o$-Ps is confined in an infinite potential well and its state is presented by the standing wave as depicted in figure \ref{fig1}(a). However, once the pore size is increased, its radius moves toward the semiclassical region \cite{Gidley99}, hence the $o$-Ps energy is in the range of $kT$ at room temperature with $k$ being the Boltzmann constant. As a result, the $o$-Ps with lower energy will exist with a longer lifetime, leading consequently to the larger pore size or the higher de-trapping probability of $o$-Ps \cite{Gow2000,Dutta08,Mahe10}. Therefore, instead of using a spherically symmetric infinite potential well ($U$ goes from 0 to infinity) as in the TE model (Fig. \ref{fig1}(a)), the SE model employed a similar potential well but with the finite depth $U = U_0$ (Fig. \ref{fig1}(b)). With this finite potential well, the assumption of standing waves as in the TE model is no longer hold. Instead, the SE model employed the Gaussian wave packets, which describe the scattering of $o$-Ps back and forth between the energy barriers on the pore wall before annihilating with electrons, similar to those proposed in Ref. \cite{Ito99}. Thus, the annihilation process of $o$-Ps will pass through both the intrinsic and pick-off annihilations in vacuum around the pore center and at the pore wall, respectively. When the pore size is increased to a large enough space, the $o$-Ps probability density function becomes mostly uniform in the pore and $\Delta R$ is very small in comparison with $R$. Therefore, the $o$-Ps wave function in the range from $R$ to $R+\Delta R$ can be approximated by $\psi_{SE}(r) = D\alpha/r$, where $\alpha$ is the normalization factor determined from the condition that the $o$-Ps probability function in the pore should be uniform and $D \approx \exp(-2\kappa\Delta R)$ ($\kappa = \sqrt{4m_e(U_0-E)/\hbar^2}$ with $m_e$ and $\hbar$ being the electron mass and Plank constant, respectively) \cite{Thanh08}. The pick-off annihilation rate of $o$-Ps in this case is then given as 
\begin{equation}
\lambda_{SE}=\lambda_A \frac{3D}{1+D} \frac{\Delta R}{R+\Delta R}\left(\frac{R}{R+\Delta R}\right)^2 = \lambda_P f_{SE}(R) ~,
\label{SE}
\end{equation}
where $\lambda_P=\frac{3D}{1+D}\lambda_A$ and $f_{SE}(R) = \frac{\Delta R}{R+\Delta R}\left(\frac{R}{R+\Delta R}\right)^2$. Equation (\ref{SE}) is formally similar to Eq. (\ref{TE}), that is, $\lambda_P$ and $f_{SE}(R)$ are equivalent to $\lambda_A$ and $f_{TE}(R)$, respectively. However, $\lambda_P$ is now expressed in terms of the diffusion probability $D$, whereas $f_{SE}(R)$  does not contain the sine function because of its simple wave function ($\psi_{SE}(r) \sim 1/r$) within the region of $R$ and $R+\Delta R$. Moreover, it should be noted here that the parameter $D$ was defined in Ref. \cite{Thanh08} as the $o$-Ps diffusion coefficient, which was then approximated by $D \approx \exp(-2)$. This artificial definition and approximation of $D$ are still difficult to physically justify. As discussed later, $D$ should be correctly considered as the probability for which the $o$-Ps could diffuse into the virtual electron layer $\Delta R$ (diffusion probability). Hence, the value of $D$ should be found via the fitting to the experimental data in a wide and diversified range of $R$, which corresponds to different geometries of the pore. 

\subsection{The hybrid model}
The SE model in Eq. (\ref{SE}) was proposed to treat the annihilation of $o$-Ps in the region of large pore $R > 1$ nm. To have a continuous estimation for all the pores, Ref. \cite{Thanh08} has proposed a semi-empirical equation (see Eq. (9) therein), which is simply a arithmetic mean of the weighted TE and SE pick-off annihilation rates, namely 
\begin{equation}
\lambda_{semi-emp} = \frac{1}{2}\bigg[(\sqrt{\lambda_{TE}\lambda_{SE}}+\lambda_T) + \sqrt{(\lambda_{TE}+\lambda_T)(\lambda_{SE}+\lambda_T)}\bigg] ~, \label{semi}
\end{equation}
In fact, the annihilation rate must closely relate to the annihilation probability and thus, the simple arithmetic mean as equation (\ref{semi}) should not be correct. Therefore, we hereby propose a correct combination of the two models, called the hybrid (HYB) model, namely 
\begin{equation}
\lambda_{HYB} = \bigg[(\sqrt{\lambda_{TE}\lambda_{SE}}+\lambda_T)\times\sqrt{(\lambda_{TE}+\lambda_T)(\lambda_{SE}+\lambda_T)}\bigg]^{1/2} ~, \label{HYB}
\end{equation}
which can be approximately shorten to 
\begin{equation}
\lambda_{HYB} \approx \sqrt{(2\lambda_{TE}+\lambda_T)(\lambda_{SE}/2+\lambda_T)} ~. 
\end{equation}
Two Eqs. (\ref{semi}) and (\ref{HYB}) produce similar results if $\lambda_T$ is relatively small compared to $\lambda_{SE}$ and $\lambda_{TE}$. 

\section{Results and discussions}
\begin{figure}
\begin{center}
      \includegraphics[scale=0.35]{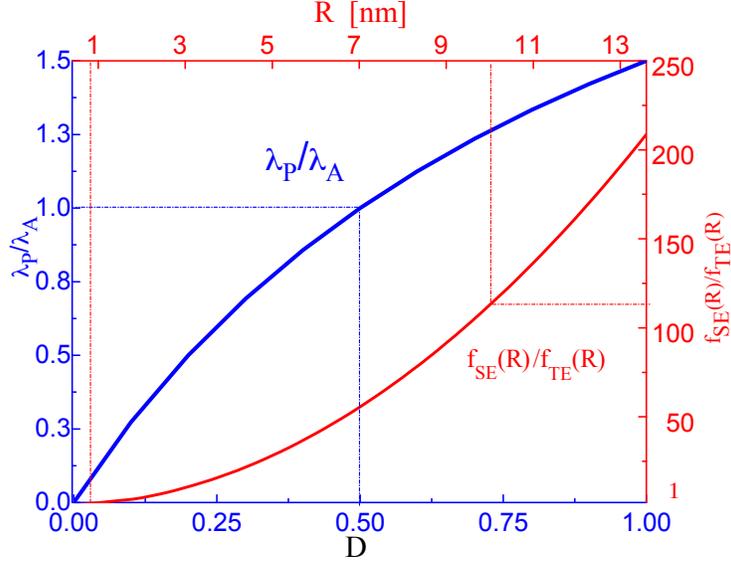}
       \caption{The plots of the relation between $\lambda_A$ and $\lambda_P$ as well as $f_{TE}(R)$ and $f_{SE}(R)$ obtained from equations (\ref{TE}) and (\ref{SE}).
        \label{fig2}}
\end{center}
\end{figure}
\begin{figure}
\begin{center}
       \includegraphics[scale=0.5]{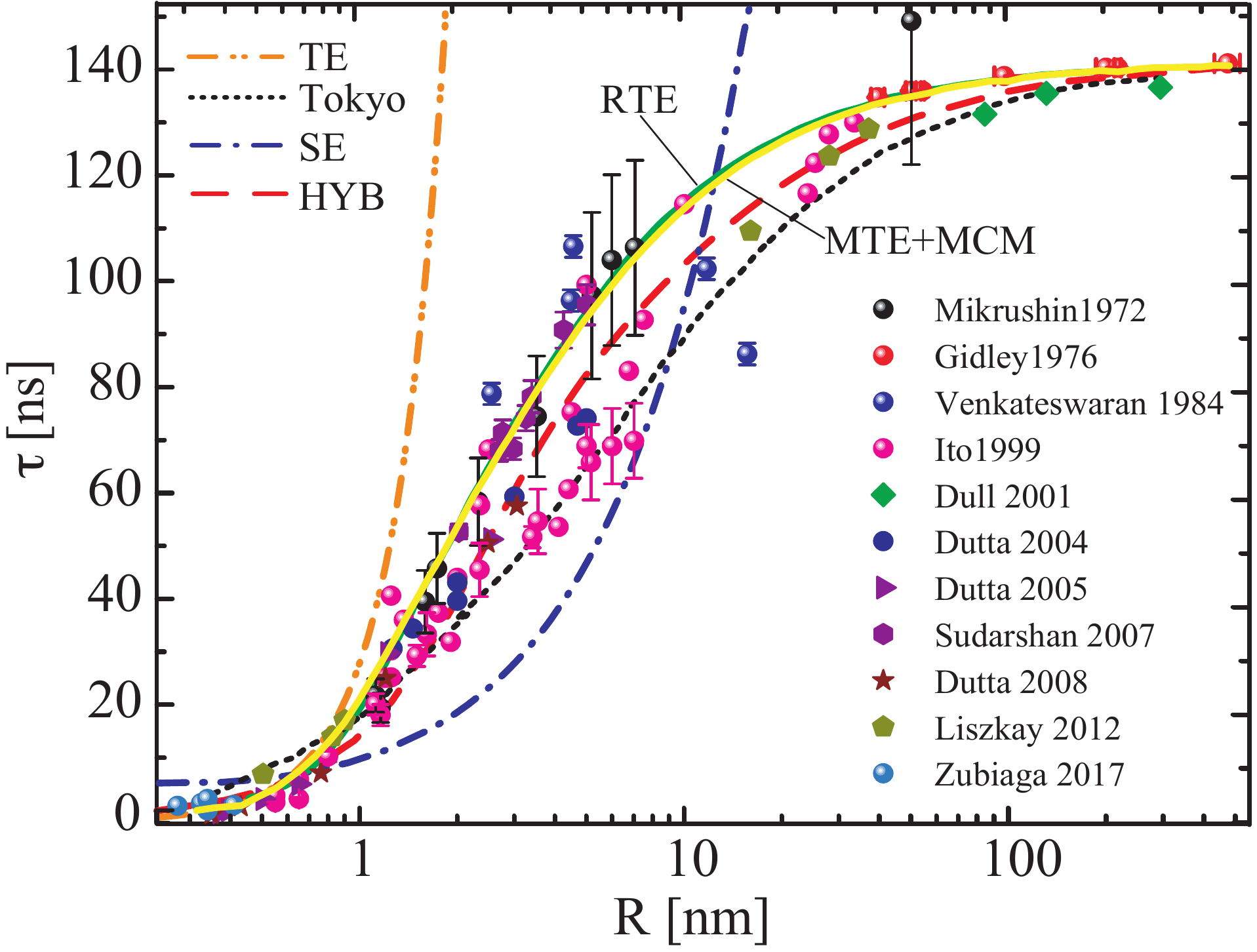}
       \caption{The $o$-Ps lifetimes $\tau$ obtained within the TE, SE, Tokyo, RTE, and HYB models versus experimental data for different porous materials. Within the HYB model, the diffusion probability $D$ is found to be 0.0985.
        \label{fig3}}
\end{center}
\end{figure}

Figure \ref{fig2} plots the relation between the ratios $\lambda_P/\lambda_A$ and $f_{SE}/f_{TE}$ versus the pore radius $R$ and diffusion probability $D$, whereas Fig. \ref{fig3} depicts the $o$-Ps lifetime $\tau$ obtained within different models (TE, SE, RTE, and HYB) in comparison with the experimental data collected from all the experiments over four decades \cite{Gidley76,Dull01,Ito99,Dutta08,Mikrushin72,Ito99a,Dutta04,Dutta05,Sudarshan07,Liszkay12,Zubiaga17}. In Fig. \ref{fig2}, the ratio $\lambda_P/\lambda_A$ increases with increasing $D$, equals to 1 at $D = 0.5$, and reaches 1.5 at the maximum value of $D=1$. In addition, $\lambda_P$ is affected by not only the pore radius $R$ but also the diffusion probability $D$. This result suggests that for the small pores, the $o$-Ps will interact and diffuse strongly to the virtual electron layer $\Delta R$, leading to the increase of the pick-off annihilation rate and the decrease of the $o$-Ps lifetime as predicted by the TE model. However, when the pore size is large enough, the $o$-Ps has some possibilities to diffuse to the virtual electron layer at the wall, resulting in the pick-off annihilation process via the emission of 2$\gamma$ or the scattering back to the center of the pore before the appearance of the 3$\gamma$ intrinsic annihilation phenomenon. Thus, the SE model reconfirms the important contribution of the intrinsic annihilation rate of $o$-Ps, which was mentioned within the Tokyo model (see Eq. (\ref{Tokyo})). As for the ratio between the two pore-size correlation functions $f_{SE}$/$f_{TE}$, Fig. \ref{fig2} shows that at $R \leq$ 0.7 nm, this ratio increases very slowly, leading to the value of $\tau_{SE}$ being lower than that of $\tau_{TE}$. With increasing $R >$ 0.7 nm, this ratio rapidly increases by more than two orders of magnitude at $R =$ 10 nm. This is a striking result, which indicates that the $o$-Ps lifetime obtained within the SE model will reach the extreme value much slower than that obtained within the TE model and thus the calculating limitation of the SE model will be more extensive in comparison with the TE model. In other words, the TE model is only applicable for the small pores, whereas the SE model is favorable for the large pores. Hence, the combination of the TE and SE models as presented in Eq. (\ref{HYB}) should result in a hybrid model, which smoothly connects the two regions of the pores and is consequently workable for all the materials with any pore sizes. This can be clearly seen via Fig. \ref{fig3}. 

In Fig. \ref{fig3}, one can see that in the small pore region ($R <$ 0.7 nm), $\tau_{TE}$ and $\tau_{RTE}$ excellently agree with the experimental data, whereas $\tau_{SE}$ and $\tau_{Tokyo}$ overestimate the data. At higher $R >$ 0.7 nm, $\tau_{TE}$ rapidly increases, while the increase of $\tau_{SE}$ is relatively slower. Also in this region, $\tau_{RTE}$ agrees with the upper data points, whereas $\tau_{Tokyo}$ fits to the lower data ones. Obviously, the HYB model (with the best fitted value of the diffusion probability $D$ is found to be 0.0985), which is a combination of the TE and SE models, has successfully connected two pore-size regions, resulting in a continuous curve of $\tau_{HYB}$ in the whole region of $R$. In addition, the values of $\tau_{HYB}$, which are in the middle of the upper and lower data points, are in the best agreement with the experimental data in comparison with other models (Tokyo and RTE). This result of the HYB model reveals the important contribution of the diffusion probability $D$ proposed in equation (\ref{SE}) of the SE model. 

\begin{figure}[h]
\begin{center}
       \includegraphics[scale=0.55]{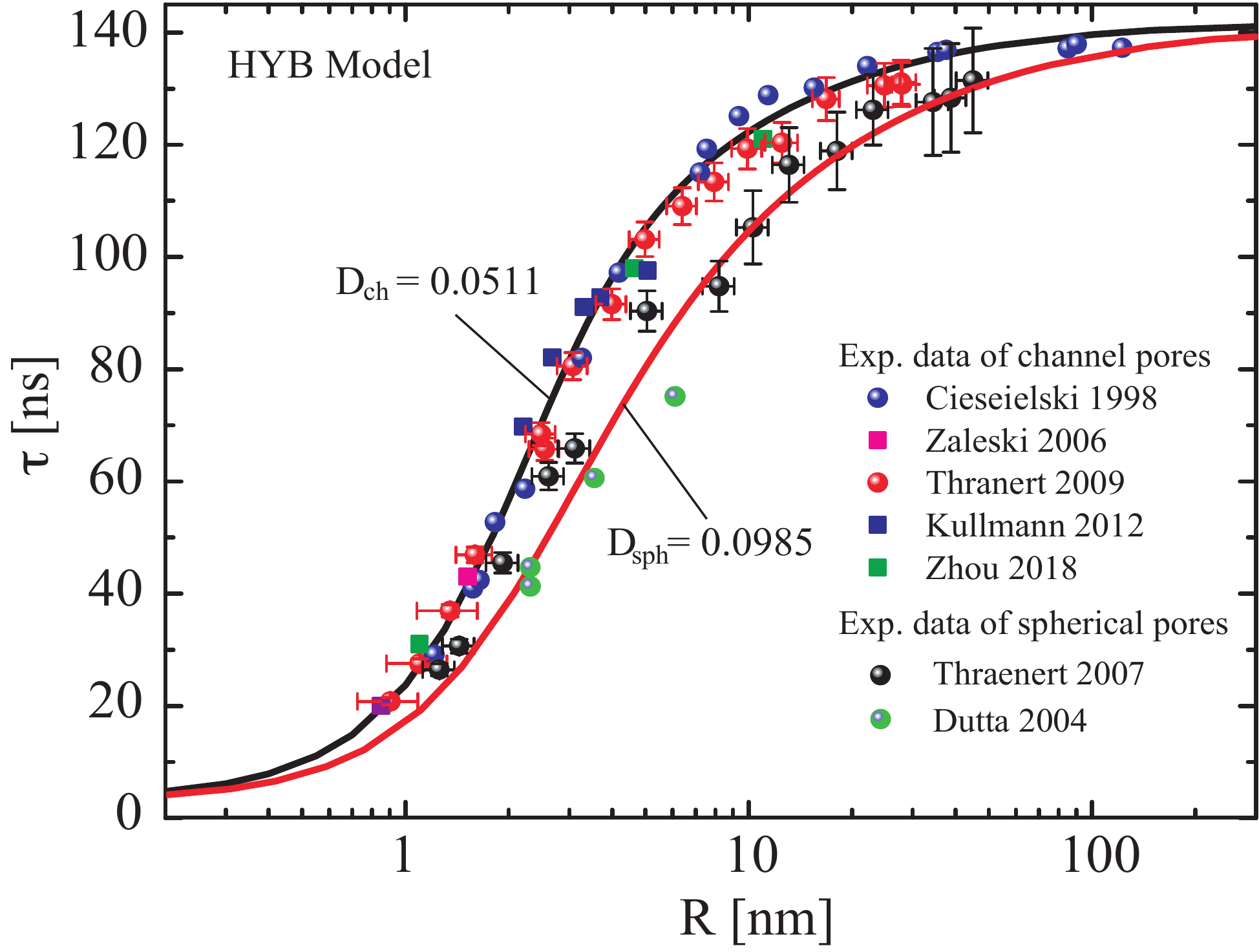}
       \caption{The $o$-Ps lifetimes obtained within the HYB models versus experimental data for the materials with spherical and channel pores.
        \label{fig4}}
\end{center}
\end{figure}

Generally, the diffusion probability $D$ should depend on the geometry of the pore. Two popular pores, which have been often studied and detected within the experiment, are those having spherical and channel geometries. Within the HYB model, the $o$-Ps lifetime also varies with $D$, so there must be different values of $D$ corresponding to different pore geometries. By selecting the experimental data detected for materials with the defined spherical \cite{Thrae07,Dutta04} and channel \cite{Ciesielski98,Zaleski06,Thrae09,Kullmann12,Zhou18} pores, we are able to select the values of $D$ so that $o$-Ps lifetimes obtained within the HYB model fit well to the corresponding experimental data. The results are shown in Fig. \ref{fig4}. Here, the values of $D$ are found to be 0.0985 ($D_{sph}$) and 0.0511 ($D_{ch}$) for the spherical and channel pores, respectively. These values of $D$ are physically reasonable because the probability that the particle diffuses into the spheres must be always higher than that occurs in the channels.

\section{Conclusion}
The present paper proposes a semi-classical (SE) model, which treats the $o$-Ps lifetime in the region of large pores where the quantum mechanical effect should be negligible in comparison with the classical one. The SE model, which takes into account the important effect coming from the diffusion probability $D$, is then consistently combined with the conventional Tao-Eldrup model for small pores, resulting a hybrid (HYB) version. As the result, the HYB model not only connects smoothly two regions of the pore but also agrees well with the experimental $o$-Ps lifetimes collected in the last four decades for the whole region of pore size from 0.2 to 400 nm. This HYB model is also a better agreement with the experimental data than the rectangular TE (RTE) and Tokyo models, which are the most up-to-date models of the $o$-Ps lifetime and pore size. In particular, by varying the diffusion probability $D$, we are able to describe very well two defined sets of experimental data for the $o$-Ps lifetime in the pores with spherical and channel geometries. The value of $D$ for the pores with the spherical geometry is found to be larger than those with the channel one. This finding is physically reasonable since the diffusion probability of the particle in the spheres must be certainly higher than that in the channels. The merit of the present model, in comparison with the previously proposed ones, is that it is applicable for the pore size in the universal range of $0.2 - 400$ nm for almost all porous materials with different geometries.

In the present model, the effect of temperature has not been considered. In fact, the temperature effect should be treated via the temperature-dependent diffusion probability $D$ since $D$ is strongly dependent on temperature. The calculations with temperature are undergoing and the results will be reported in a separate paper. 

\section*{Acknowledgements}
This work is funded by Ministry of Science and Technology of Vietnam under the Grant Number DTCB: 14/19 TTHN.

\end{document}